\documentstyle[epsf]{res}
\begin{document}

\title{%
Reionization of an Inhomogeneous Universe
}

\author{Masayuki UMEMURA, Taishi NAKAMOTO, and Hajime SUSA
\\
{\it Center for Computational Physics, \\ University of
  Tsukuba, Tsukuba 305, Japan, \\
\{umemura, nakamoto, nakamoto \}@rccp.tsukuba.ac.jp}}

\maketitle

\baselineskip=5.7mm

\noindent
{\bf Abstract}
\vspace{2mm}

   Full radiative transfer in 3D space has been solved to pursue the reionization
history in an inhomogeneous universe. It has been shown that the reionization of an
inhomogeneous universe is not a prompt event, but a fairly slow process. Also, QSO
absorption line systems are simulated with using the results of radiative transfer
calculations. Ly$\alpha$ continuum depression
implies that the metagalactic UV
intensity decreases rapidly with $z$ at $z>5$, 
and the reionization must have taken place
between $z=6$ and 10. Finally, it is stressed 
that H$\alpha$ forest is a more powerful tool to
probe the reionization history and the density fluctuations 
in the universe at $z>5$.

\section{Introduction}

   The cosmic reionization is one of the most significant issues in cosmology, 
which is closely related to the formation of QSOs and galaxies. The information 
on the ionization states in the universe has been accumulated by the observations 
of Ly$\alpha$ absorption lines in high redshift QSO or galaxy spectra. 
3D cosmological hydrodynamic simulations 
(Cen et al. 1994; Miralda-Escude et al. 1996; Gnedin \& 
Ostriker 1996; Zhang et al. 1997) 
have revealed that the Ly$\alpha$ absorption 
systems can be accounted for in terms of the absorption by intergalactic density 
fluctuations. However, all of 
these works have been based upon optically-thin or local optical-depth 
approximations. Recently, it has been shown that the radiative transfer effects of 
ionizing radiation could strongly affect the ionization structure 
(Razoumov \& Scott 1999; Gnedin 1999; Nakamoto et al. 1999). 
Madau (1995) and Haardt \& Madau (1996) 
as well have considered cosmological radiative transfer on the 
assumption of 1D semi-infinite slab. Here, we present 3D radiative transfer 
calculations on the cosmic reionization and the formation of QSO absorption line 
systems. 

\section{Method}


   To reproduce an inhomogeneous universe, we generate density fluctuations 
based upon the Zel'dovich approximation.
In the present calculations, we assume a standard 
cold dark matter cosmology, i.e., 
$\Omega_{\rm CDM}=0.95$, $\Omega_{\rm Baryon}=0.05$, 
$\sigma_8=0.6$ with the Hubble constant of 50 km s$^{-1}$ Mpc$^{-1}$. 


   We calculate the ionization degree by assuming the temperature of $10^4$K 
and ionization equilibrium. The simulation box is irradiated by the isotropic UV 
background radiation of a power law-type spectrum, 
$I_0=I_{21}10^{-21}$ erg cm$^{-2}$ s$^{-1}$ Hz$^{-1}$ sr$^{-1}$.
The so-called "proximity effect" of Lyman alpha absorption lines 
requires the diffuse UV radiation to be at a level of $I_{21} \approx 1$
at $2<z<4$ (Bajtlik, Duncan, \& Ostriker 1988; Giallongo et al. 1996). 


   The UV radiation fields are obtained by solving the 
three-dimensional steady radiative transfer equation. 
For the implementation of the radiative transfer 
on a massively parallel supercomputer, 
we have developed a new scheme 'Sequential Wave Front Method'. We have used 
$128^3$ grids in space, $128^2$ in directions, and 6 
for important line frequencies of 
hydrogen and helium. Since the continuum is analytically integrated, the 
calculations are quite accurate even if just 6 frequencies are solved. The total 
operations amounts to about 1 Tflops*hour. The calculations have been 
performed on the CP-PACS in University of Tsukuba. 

\section{Results \& Discussion}

\subsection{Cosmic Reionization}

   Around $z=15$, underdense regions are reionized and percolate, leaving 
overdense neutral islands due to the self-shielding effects. Since the absolute 
density decreases with time in a linear regime of density fluctuations, the 
reionization proceeds and the ionized sea encroaches onto the neutral islands, 
leaving filamentary self-shielded regions at $z \sim 9$. 
The highly ionized regions expand further at $z \sim 7$, 
but the universe is not perfectly transparent against UV 
radiation in the sense that the collective optical depth considerably reduces the 
incident radiation. Around $z=5$, the universe becomes transparent against 
background UV and the overall reionization has been fulfilled. The simulations 
have revealed that, owing to the radiative transfer effects, the reionization in an 
inhomogeneous universe is not a prompt event , but fairly slow process, in 
contrast to the prompt reionization in a homogeneous universe. 

\subsection{Ly$\alpha$ Absorption Lines and Continuum Depression}

   The resultant ionization degrees in the universe are different from place to 
place by more than three orders of magnitude. Due to such inhomogeneous 
ionization structure, relatively low ionization regions could produce strong 
absorption in quasar spectra. To make a direct comparison with the observations, 
we simulate absorption lines. 
First, we focus on Ly$\alpha$ absorption. 
To match the recent observations by the Keck telescope, 
we adopt the resolution of 
R=45000 and the variance of 0.04, and assume the Voigt profile of lines. 
The simulated Ly$\alpha$ absorption features are 
shown in Figure 1. 

\vspace{-3mm}
\baselineskip=3.8mm
\begin{figure}[hbtp]
\begin{flushleft}
\epsfile{file=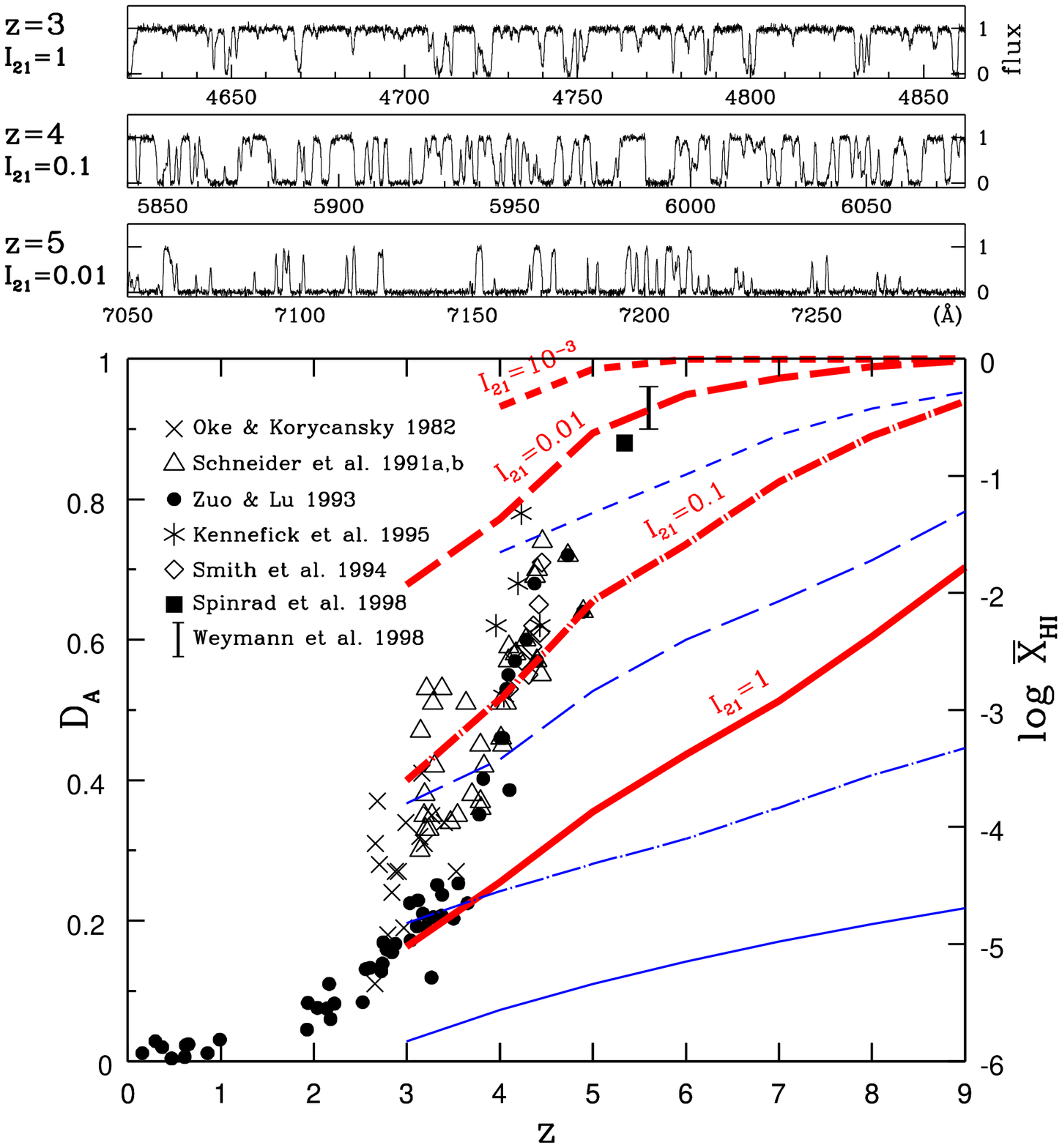,height=7cm}
\end{flushleft}
\end{figure}
\vspace*{-0.8cm}
\parshape=1 0cm 7cm
\noindent 
{\scriptsize Fig.1 --
The simulated Ly$\alpha$ absorption lines against wavelength 
at $z=3$ ({\it top panel}) with $I_{21}=1$, $z=4$ ({\it second}) 
with $I_{21}=0.1$, and $z=5$ ({\it third}) with $I_{21}=0.01$. 
The bottom panel is the diagram of 
Ly$\alpha$ continuum depression
({\it thick gray curves}) against redshifts. Symbols are
observations. 
Also, the mean neutral fractions $X_{HI}$ ({\it thin curves}) are 
shown. The same line types correspond to the same UV intensity.
}

\vspace*{-10.7cm}
\begin{figure}[hbtp]
\begin{flushright}
\epsfile{file=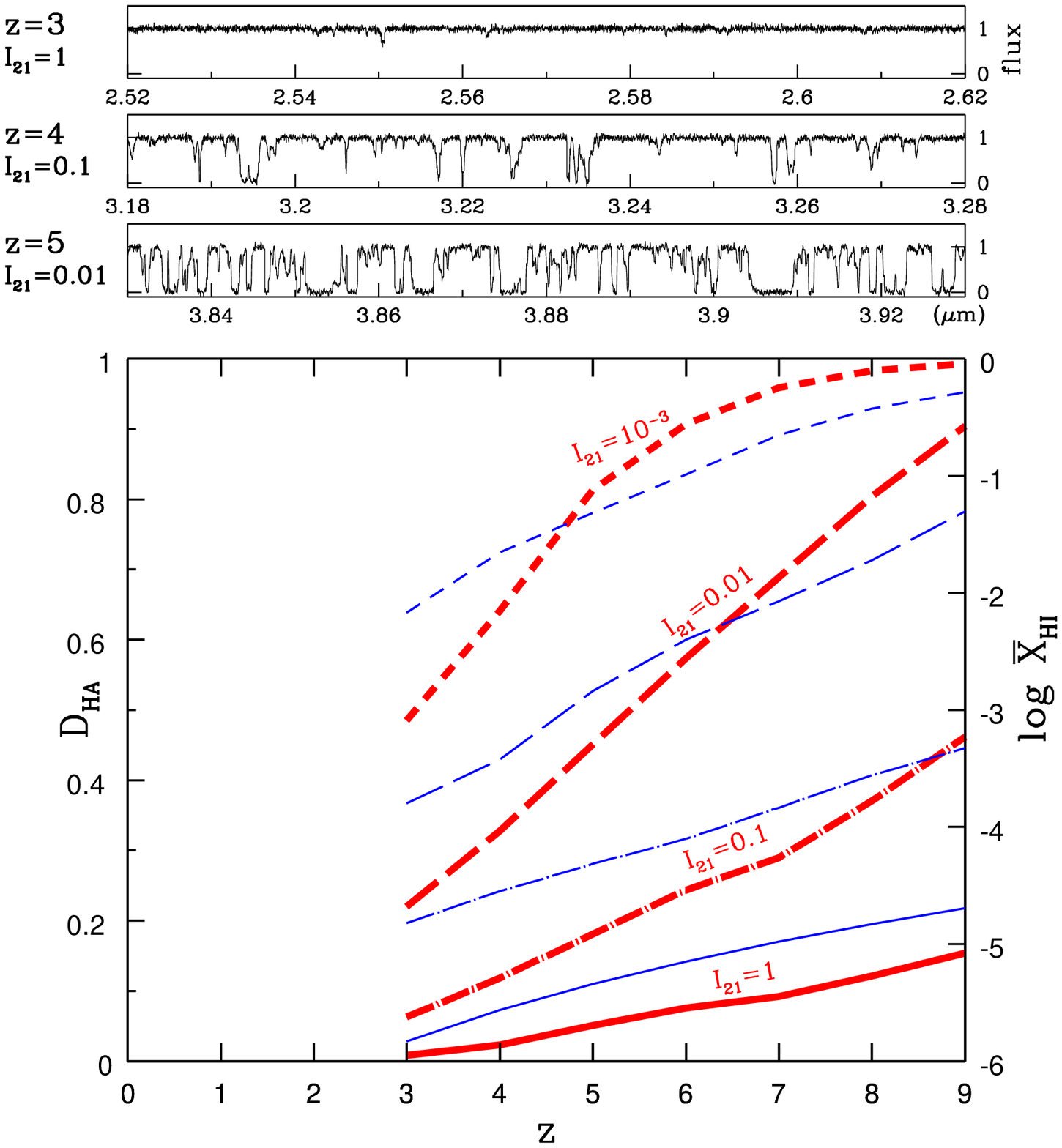,height=7cm}
\end{flushright}
\end{figure}
\vspace*{-0.8cm}
\parshape=1 7.5cm 7cm 
\noindent
{\scriptsize Fig.2 -- 
Same as Fig. 1, but for the H$\alpha$ absorption. 
There are no observation data so far for H$\alpha$ absorption at high redshifts
}

\vspace*{2.1cm}

\baselineskip=5.5mm

In order to 
compare quantitatively the simulations with observations, we have assessed the 
so-called continuum depression, $D_A$.
Figure 1 shows that any model with a constant UV intensity does not match 
the observed trend that $D_A$ tends to grow quickly at higher redshifts
up to 5.
This implies that the metagalactic UV intensity must decrease 
rapidly with $z$ at $z>5$ by two orders of magnitude at least. 
If the well fitted value of $I_{21} =0.01$ at $z=5$ is 
unchanged also at higher redshifts, the reionization epoch is estimated 
to be $z \approx 9$.  
If the UV intensity decreases in a similar fashion also at higher redshifts, the 
reionization epoch is $z \approx 6$. 
Thus, it is concluded that the cosmic reionization must 
have taken place between $z=6$ and 10. However, the Ly$\alpha$ absorption is not 
appropriate to determine the reionization epoch more accurately, because, as seen 
in the absorption features in Figure 1, Ly$\alpha$ is too strongly 
depleted even if the 
mean neutral fraction is less than $10^{-2}$. 
In other words, $D_A$ is no longer sensitive to 
$X_{HI}$ around the cosmic reionization epoch.

\subsection{H$\alpha$ Forest}

   As shown in the previous subsection, Ly$\alpha$ has too high line 
opacity to probe the universe at $z>5$. Therefore, three conditions are 
required for a line in order to 
investigate the universe at $z>5$: 
(1) it has lower line opacity than Ly$\alpha$, 
(2) line emission is detectable, and 
(3) it has lower extinction against dust because young 
star-forming galaxies are often dust-enshrouded. The most favorable solution 
is H$\alpha$ absorption lines. In Figure 2, the H$\alpha$ absorption features 
and continuum depression are shown. 
The H$\alpha$ absorption is relatively weak around $z=3$, 
while it is sensitive to the ionization degrees at $z>4$. 
Also the H$\alpha$ continuum depression 
traces the reionization history more accurately. Therefore, it is concluded that 
H$\alpha$ forest is a more powerful tool to probe the universe at $z>5$. 
H$\alpha$ forest has been never detected so far. The reason comes from the fact that 
H$\alpha$ has much weaker opacity than Ly$\alpha$. 
From observational points of view, the 
continuum depression can be detected by low-dispersion spectroscopy 
or narrow-band photometry. Furthermore, H$\alpha$ forest is subject 
to less UV bump effects for 
AGNs compared to Ly$\alpha$ forest. The wavelengths of H$\alpha$ forest drop on 
$3\mu {\rm m} \lsim \lambda_{{\rm H}\alpha} \lsim 7 \mu {\rm m}$ 
at $4\lsim z \lsim 10$. 
Thus, the observations can be done with Subaru IRCS, 
IRIS, SIRTF, NGST, or H2/L2. If one can obtain the absorption features with the 
resolution greater than 10000, one can recover the density fluctuations at high 
redshifts. They allow us to determine the linear amplitude of pregalactic 
perturbations which is by no means measured in the CBR due to the strong 
Sunyaev-Zeldovich effects in galactic scales. If one has the amplitude of linear 
density fluctuations at galactic scales, one can not only set 
the initial condition for 
galaxy formation, but also make more reliable determination of cosmological 
parameters.

\vspace{1pc}

\small

\noindent
{\bf References}
\vspace{2mm}
\re
Bajtlik, S. Duncan, R. C., Ostriker, J. P. 1988, ApJ, 327, 570
\re
Cen, R., Miralda-Escude, J., Ostriker, J., Rauch, M. 1994, ApJ, 437, L9
\re
Giallongo, E., Cristiani, S., D'Odorico, S., Fontana, A., Savaglio, S. 1996, ApJ, 466, 46
\re
Gnedin, N. 1999, astro-ph/9909383 
\re
Haardt, F., Madau, P. 1996, ApJ, 461, 20 
\re
Madau, P. 1995, ApJ, 441, 18
\re
Miralda-Escude, J, Cen, R., Ostriker, J. P., Rauch, M. 1996, ApJ, 471, 582
\re
Nakamoto, T., Umemura, M., Susa, H. 1999, MNRAS, submitted
\re
Razoumov, A. O., Scott, D. 1999, MNRAS, 309, 287
\re
Zhang, Yu, Anninos, P., Norman, M. L., Meiksin, A. 1997, 485, 496

\end{document}